\documentclass[runningheads]{svmult}
\usepackage{makeidx}   % allows index generation
\usepackage{graphicx}  % standard LaTeX graphics tool
\usepackage{subeqnar}  % subnumbers individual equations
\usepackage{multicol}  % used for the two-column index
\usepackage{cropmark} % cropmarks for pages without
\usepackage{physprbb}  % centered layout of diverse elements, etc.
\input epsf.tex
\def\DESepsf(#1 width #2){\epsfxsize=#2 \epsfbox{#1}}
\begin{document}
\title*{DARK MATTER IN SUPERGRAVITY}
\toctitle{DARK MATTER IN SUPERGRAVITY}
% allows explicit linebreak for the table of content
%
%
%\titlerunning{Focusing of a Parallel Beam}
% allows abbreviation of title, if the full title is too long
% to fit in the running head
%
\author{R. Arnowitt
\and B. Dutta
\and Y. Santoso}

\institute{Department of Physics, Center for Theoretical Physics, 
Texas A\&M University, College Station, TX 77843-4242, USA}

\maketitle              % typesets the title of the contribution

\begin{abstract}We consider neutralino-proton cross sections for halo dark matter
neutralinos ($\tilde\chi_1^0$) within the framework of supergravity models with
R-parity invariance for models with universal soft breaking (mSUGRA) and
models with nonuniversal soft breaking. The analysis includes the necessary
corrections to treat the large tan$\beta$ region (i.e. L-R mixing in the
squark and slepton mass matrices, loop corrections to the $b$ and $\tau$
masses,etc) and includes all coannihilation phenomena. For mSUGRA, dark
matter detectors with current sensitivity are seen to be probing the region
where tan$\beta \stackrel{>}{\sim}$25, $\Omega_{\tilde\chi_1^0} h^2 <0.1$, 
$m_{\tilde\chi_1^0}\stackrel{<}{\sim}90$ GeV, and for the light Higgs,
$m_h \stackrel{<}{\sim}120$ GeV. Nonuniversal models can have a much larger cross section, and
current detectors can probe part of the parameter space where 
tan$\beta\stackrel{>}{\sim}4$.
Minimum cross sections are generally greater than $10^{-9}$ pb to $10^{-10}$ pb
for $m_{1/2} <600$ GeV (and hence accessible to planned future detectors), with
the exception of a region when $\mu <0$ where for 
$m_{1/2}\stackrel{>}{\sim}450$ GeV, $4\stackrel{<}{\sim} \tan\beta
\stackrel{<}{\sim}20$, the cross section drops to a minimum of about 
$1\times10^{-12}$ pb at
$m_{1/2}=600$ GeV, $\tan \beta \simeq10$. In this region, the gluino and squarks lie
above 1 TeV, but should still be accessible to the LHC.\end{abstract}

\section{Introduction}

If the dark matter that exists in the Milky Way is a supersymmetric weakly
interacting particle (wimp), there are several ways in which it might be
detected. Annihilation of two wimps in the halo might give rise to signals
of gamma rays, anti-protons or positrons. Dark matter particles caught by
the gravitational fields of the Sun or Earth would be expected to sink to
the center, and there annihilate leading to neutrinos that might be
detected on the surface of the earth. Finally direct detection of incident
wimps from their scattering by nuclear targets on the Earth is possible. Of
these, the last possibility appears most promising, and there are now a
large number of detectors searching for supersymmetric wimps. We consider
here what signals might be available within the framework of
supergravity(SUGRA) models with grand unification of the gauge coupling
constants at the GUT scale $M_G\cong2\times10^{16}$GeV.
There are three different types of SUGRA models currently being
investigated, which differ by the mechanisms used to achieve
supersymmetry(SUSY) breaking in the physical sector. These are gravity
mediated SUGRA models, gauge mediated models, and anomaly mediated models.
Of these, the gravity mediated models with R-parity invariance have the
most robust candidate for particle dark matter, and we will restrict our
discussion here to such models. In gravity mediated models, the dark matter
particle is the lightest supersymmetric particle, the LSP, (absolutely
stable due to the R-parity invariance), and this is generally the lightest
neutralino, $\tilde\chi_1^0$. The nucleus-$\tilde\chi_1^0$ scattering cross section contains a spin
independent part and a spin dependent part. For heavy nuclear targets, the
spin independent scattering dominates, and it is possible to extract from
data the $\tilde\chi_1^0$-proton cross section, $\sigma_{\tilde\chi_1^0-p}$. There are a number of
astronomical uncertainties, but making conventional assumptions, current
detectors (DAMA, CDMS, UKCDM) are sensitive to halo $\tilde\chi_1^0$ if
\begin{equation}
             \sigma_{\tilde\chi_1^0-p} \stackrel{>}{\sim} 1\times10^{-6} {\rm pb}                                
\end{equation}
and future detectors (GENIUS, Cryoarray) plan to achieve sensitivities of
\begin{equation}
             \sigma_{\tilde\chi_1^0-p} \stackrel{>}{\sim} (10^{-9} -10^{-10}){\rm pb}                                
\end{equation}

\begin{figure}[htb]
\centerline{ \DESepsf(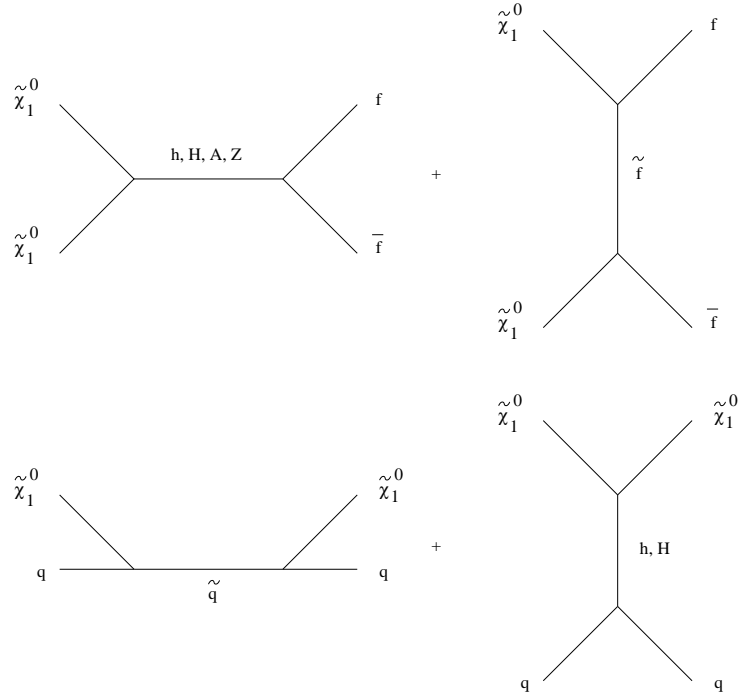 width 10 cm) }
\caption {\label{fig1} Dominant diagrams for $\sigma_{\rm ann}$ (upperdiagrams) and spin
independent part of $\sigma_{\tilde\chi^0_1-p}$(lower diagrams).}
\end{figure}

We discuss here how these sensitivities might relate to supergravity
models. In particular, we consider the minimal supergravity GUT model
(mSUGRA)[1] which has universal soft breaking masses at $M_G$, and
nonuniversal soft breaking models[2] which allow nonuniversal Higgs masses
and nonuniversal third generation squark and slepton masses at $M_G$ (but keep
the gaugino masses universal at $M_G$). While the models are physically
different, they lead to qualitatively similar results: Current detectors
are sensitive to a significant part of the SUSY parameter space, and future
detectors should be able to cover all of the parameter space, except for
special regions where there is an accidental cancelation of terms making
$\sigma_{\tilde\chi^0_1-p}$ anomalously small.
Each of the above models contains a number of arbitrary new parameters. In
spite of this they can still make relevant predictions for two main
reasons: (i) Using the renormalization group equations (RGE) starting from
$M_G$, they allow for radiative breaking of $SU(2)\times U(1)$ at the electroweak
scale (and thus furnish a natural explanation for the Higgs mechanism);
(ii) Along with being able to calculate $\sigma_{\tilde\chi^0_1-p}$, the models can also
calculate the relic density of neutralinos, i.e. $\Omega_{\tilde\chi_1^0} h^2$, where 
$\Omega_{\tilde\chi_1^0}
= \rho_{\tilde\chi_1^0}/\rho_c$, $\rho_{\tilde\chi_1^0}$ is the relic mass density of the 
${\chi_1^0}$, and
$\rho_c=3H_0/8\pi G_N$. Here $H_0=h(100\mbox{km/s Mpc})$ is the Hubble constant, and $G_N$ is
the Newton constant. Both of the above lead to important constraints on the
SUSY parameter space. Thus one has that
\begin{equation}
         \Omega_{\tilde\chi^0_1}h^2\sim(\int^{x_f}_0 dx\langle\sigma_{\rm
ann}v\rangle)^{-1}
\end{equation}
where $\sigma_{\rm ann}$ is the annihilation cross section in the early universe, v
is the relative  neutralino velocity at annihilation, and $<...>$ means
thermal average. The dominant Feynman diagrams for $\sigma_{\rm ann}$ and spin
independent $\sigma_{\tilde\chi^0_1-p}$ are shown in
Fig.\ref{fig1}, and roughly speaking $\sigma_{\rm ann}$ depends on the crossed diagrams
relative to $\sigma_{\tilde\chi^0_1-p}$. Thus usually, then, when $\sigma_{\tilde\chi^0_1-p}$ is large,
$\sigma_{\rm ann}$ will also be large, and hence by Eq(3), $\Omega_{\tilde\chi_1^0}h^2$ will be small.
Thus lower bounds on $\Omega_{\tilde\chi_1^0}h^2$ will produce upper bounds on 
$\sigma_{\tilde\chi^0_1-p}$.
In the following, we will assume $h=0.70\pm0.07$ and for matter(m) and
baryonic matter(b) the values $\Omega_m=0.3\pm0.1$, and $\Omega_b=0.04$. (This
corresponds to a dark energy amount of $\Omega_\Lambda \sim0.65$). For the dark
matter then one has $\Omega_{\tilde\chi_1^0}=0.26\pm 0.10$, and if one combines errors in
quadrature, one finds $\Omega_{\tilde\chi_1^0} h^2 =0.13\pm0.05$. Since there is undoubtedly a
large amount of systematic error in the above estimates, in the following
we will assume the approximately 2$\sigma$ spread of
\begin{equation}
                 0.02 < \Omega_{\tilde\chi_1^0} h^2 < 0.25                                
\end{equation}
The lower bound of Eq.(4) lies somewhat below other estimates. However, it
also allows for the possibility that not all the dark matter in the Galaxy
are neutralinos (e.g. some may be machos).
In addition to the above, there are accelerator bounds that constrain the
SUSY parameter space. In the following we use the LEP bounds for the light
Higgs ($h$) of $m_h > 104$ GeV for tan$\beta$ = 3, $m_h > 102$ GeV for tan$\beta=$5 and for
the light chargino $ m_{\tilde\chi_1^\pm} > 102$ GeV. (For tan$\beta > $5, the Higgs mass
bounds do not restrict the parameter space significantly.) The Tevatron
gives the gluino ($\tilde g$) mass bound of $m_{\tilde g} > 270$ GeV (for gluino and squarks
nearly degenerate). In addition there is the CLEO measurment of the
$b\rightarrow s +
\gamma$ decay. We take here a 2$\sigma$ range around the experimental central
value of the $b\rightarrow  X_s + \gamma$ branching ratio[3]:
 \begin{equation} 
 1.8 \times10^{-4} < B(B\rightarrow X_s \gamma) < 4.5 \times10^{-4}                    
\end{equation}
Of the above, the most significant constraints come from the Higgs mass
bounds and the $b \rightarrow s\gamma$ branching ratio.
\section{ Theoretical Analysis}

In order to get accurate results, it is necessary to include a number of
corrections in the calculations. We list some of these here: (1) One needs
to run the two loop gauge RGE and one loop Yukawa RGE from $M_G=2\times10^{16}$
GeV 
down to the electroweak scale, iterating to get a consistent SUSY mass
spectrum. (2) Below the SUSY scale $M_S$, one runs the QCD RGE for
contributions dominated by light quarks. (3) It turns out that results are
somewhat sensitive to bounds on the Higgs mass $m_h$, and so one needs to use
the one loop, two loop and pole mass corrections to accurately calculate
the value of $m_h$. (4) L-R mixing in the sfermion mass matrices must be
included. These are important for large tan$\beta$ and in the third generation.
(5) One loop corrections to $m_b$ and $m_\tau$ are included. This is needed to
get the correct value of the $b$ and $\tau$ Yukawa coupling constants, and again
are important for large $\tan \beta$. (6) Leading order (LO)[4] and some next to
leading order (NLO)[5] corrections to the $b \rightarrow s\gamma$ decay are included.
All of the above are under good theoretical controll except perhaps for the
$b \rightarrow s\gamma$ constraint\footnote{Recent analyses[6] appear to have calculated the most important NLO
corrections to the branching ratio for $b \rightarrow s\gamma$ for large tan$\beta$. These
corrections have not been treated here, but will be included in [7]  (where
the bounds on $m_h$ will also be updated). We do not believe this will effect
the predictions of the maximum and minimum cross sections given below, but
may modify which regions of parameter space get excluded.}.
We note that we do not make any assumptions on the nature of the GUT group
at grand unification. Hence we do not impose $b-\tau$ (or $ b-\tau-t$) Yukawa
unification at $M_G$, and do not impose proton decay constraints. Such
phenomena depend sensitively on unknown post-GUT physics, and so the
validity of these constraints are unclear. For example, string models in
which there is Wilson line breaking of the GUT group to the Standard Model
group at $M_G$, require gauge coupling constant unification but neither Yukawa
unifications implied by the GUT group nor the SUGRA proton decay
constraints need hold[8].

SUSY theory allows one to calculate neutralino-quark scattering (Fig.\ref{fig1}),
and one must convert this to neutralino-proton scattering to compare with
experiment. To do this we follow the proceedures of [9], which requires
three parameters: (i) the pion-nucleon sigma term, $\sigma_{\pi N}= 1/2(m_u +
m_d)<p|u{\bar u} +d{\bar d}|p>$, (ii) 
$\sigma_0 = \sigma_{\pi N} - (m_u + m_d)<p|s
{\bar s}|p>$,
and (iii) the quark mass ratio $r=2m_s/(m_u + m_d)$. We use here the values
$\sigma_{\pi N} = 65$ MeV (based on analyses[10] making use of recent $\pi-N$
scattering data), $\sigma_0 = 30$ MeV [11], and $r=24.4\pm1.5$[12]. If one were
to use instead the value $\sigma_{\pi N}=45$ MeV (based on older $\pi-N$ data) then
the value of $\sigma_{\tilde\chi^0_1}-p$ would be reduced by a factor of about 3.

\section{ mSUGRA Model}

\subsection{Introduction}

The mSUGRA model has universal soft breaking and so depends on a minimum
number of new parameters i.e. four parameters and one sign. These are (1)
$m_0$, the universal scalar particle mass at $M_G$. (2) $m_{1/2}$, the universal
gaugino mass at $M_G$. (Alternately, one may use 
$m_{\tilde\chi^0_1} $ or $m_{\tilde g}$ since these scale
approximately with $m_{1/2}$, 
i.e. $m_{\tilde\chi^0_1} \simeq 0.4m_{1/2}$, and m$_{\tilde g} \simeq 2.8 m_{1/2})$. (3)
$A_0$, the universal cubic soft breaking mass at $M_G$. (4) $\tan \beta =
<H_2>/<H_1>$, where $<H_2>$ gives mass to the $u$-quarks and $<H_1>$ gives mass to
the $d$-quarks and charged leptons. In addition, the sign of the Higgs mixing
parmeter $\mu$ is undetermined. ($\mu$ appears in the superpotential $W$ as
$\mu H_1H_2$.)
We take for this parameter space the following ranges:
\begin{equation}
            m_0  \leq \,1 \mbox{ TeV}
\end{equation}
\begin{equation}
            m_{1/2} \leq 600 \mbox{ GeV}\,\, 
	    \mbox{(which corresponds to } m_{\tilde g}\leq1.5 \mbox{ TeV},\,\,
	     m_{\tilde\chi^0_1} \leq
240\, \mbox{ GeV)}\end{equation}
\begin{equation}
            2 < \tan\beta < 50
\end{equation}
\begin{equation}
            |A_0/m_0| \leq5
\end{equation}

If one increases the $m_{1/2}$ bound to $m_{1/2}=$ 1 TeV (corresponding to 
$m_{\tilde g}$
=2.5 TeV, which is the upper detection limit for gluinos at the LHC), the
neutralino-proton cross section will drop by a factor of about 2-3 at the
high end of the parameter space.

\subsection{ Maximum Cross Section}

\begin{figure}[htb]
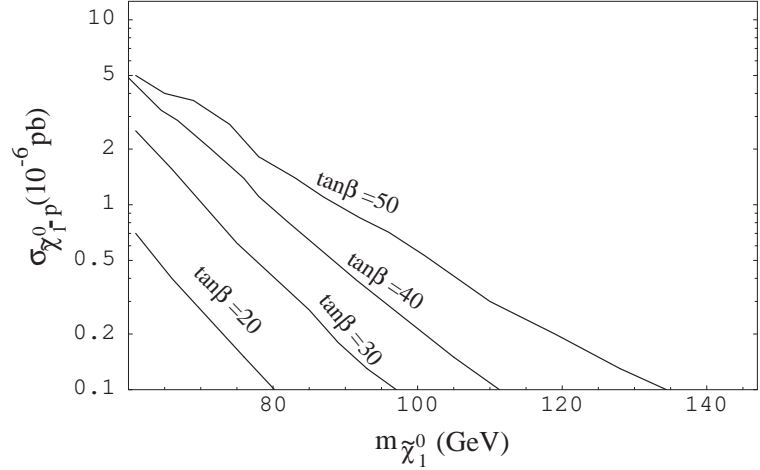

\centerline{ \DESepsf(aads20304050.epsf width 10 cm) }
\caption {\label{fig2}  Maximum $\sigma_{\tilde\chi^0_1-p}$ as a function of 
$m_{\tilde\chi^0_1}$ for tan$\beta=$ 20,30,40,50,
obtained by varying $A_0$ and $m_0$ over the parameter space[13]. The constraint on
$\Omega_{\tilde\chi^0_1}h^2$ of Eq.(4) has been imposed.}
\end{figure}

\begin{figure}[htb]
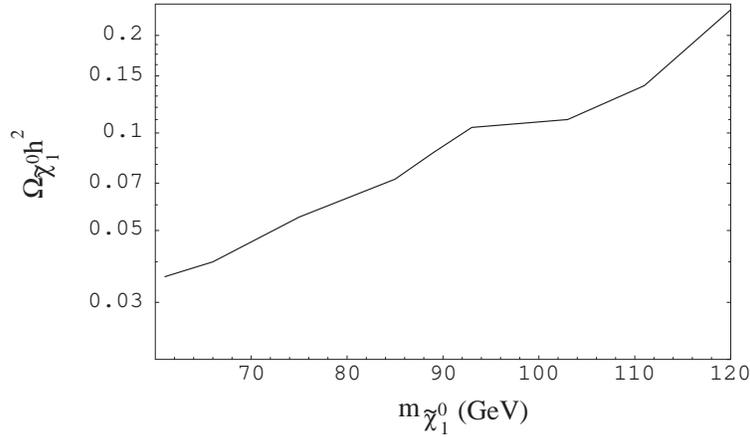

\centerline{ \DESepsf(aads30om.epsf width 10 cm) }
\caption {\label{fig3} $\Omega_{\tilde\chi^0_1}h^2$ as a function of $m_{\tilde\chi^0_1}$
 for tan$\beta=$30 when $\sigma_{\tilde\chi^0_1-p}$ takes on
its maximum value (as in Fig.\ref{fig2})[13].}
\end{figure}

We examine first the maximum cross section the model can achieve. 
$\sigma_{\tilde\chi^0_1-p}$
is an increasing function of tan$\beta$, and a decreasing function of m$_{1/2}$ and
$m_0$. Thus the maximum $\sigma_{\tilde\chi^0_1-p}$ should occur at large tan$\beta$ and small 
$m_{\tilde\chi^0_1}$.
This is illustrated in Fig.\ref{fig2} where the maximum cross section is plotted at
a function of $m_{\tilde\chi^0_1}$ for tan$\beta$ = 20, 30, 40, 50, in the range of cross sections
acessible to current detectors. One sees that current detectors with the
sensitivity of Eq.(1), have begun to sample part of the parameter space for
tan$\beta\stackrel{>}{\sim}$ 25. Further, from the
maximum tan$\beta=$ 50 curve, one sees that only neutralinos with mass 
$m_{\tilde\chi^0_1}\leq$
90 GeV are accessible to such detectors.
Fig.\ref{fig3} shows $\Omega_{\tilde\chi^0_1-p}h^2$ as a function of 
$m_{\tilde\chi^0_1}$ for $\tan\beta =30$, when $\sigma_{\tilde\chi^0_1-p}$
takes on its maximum value of Fig.\ref{fig2}. One sees that 
$\Omega_{\tilde\chi^0_1}h^2$ is an
increasing function
of $m_{\tilde\chi^0_1}$ as expected from the discussion in Sec.1, i.e. since 
$\sigma_{\tilde\chi^0_1-p}$
decreases with $m_{\tilde\chi^0_1}$, one expects that the early universe annihilation cross
will similarly decrease, and hence $\Omega_{\tilde\chi^0_1}h^2$ will increase by Eq.(3).
Since $m_{\tilde\chi^0_1}\leq$ 90 GeV for current detector sensitivities, we see that current
detectors are accessing only the region where 
$\Omega_{\tilde\chi^0_1}h^2\leq 0.1$. It is
clear that the future very accurate determinations of $\Omega_{\rm CDM}$ by the MAP
and Planck satellites will greatly sharpen the predictions of the SUGRA
models.
Fig.\ref{fig4} shows the light Higgs mass for tan$\beta=$30 when $\sigma_{\tilde\chi^0_1-p}$ 
takes on its
maximum value. For $m_{\tilde\chi^0_1}< $90GeV (the range accesible by current detectors) one
has
\begin{figure}[htb]
\centerline{ \DESepsf(aads30higgs.epsf width 10 cm) }
\caption {\label{fig4} m$_h$ as a function of $m_{\tilde\chi^0_1}$ for tan$\beta=$30, 
when $\sigma_{\tilde\chi^0_1-p}$ takes on it
maximum value[13].}
\end{figure}
$m_h\leq$120 GeV. Such a range of Higgs mass would be accessible to 
RUN2 at the
Tevatron, if the run achieves maximum luminosity.

\subsection{  Minimum Cross Sections}

We turn next to consider how small the $\sigma_{\tilde\chi^0_1-p}$ cross sections can get to see
how sensitive future detectors must be to cover the full parameter space.
It is convenient to divide the discussion into the region below
coannihilation effects and the region where coannihilation can take place.

\subsubsection{Below coannihilation ($m_{\tilde\chi^0_1}\leq150$ GeV)}

\begin{figure}[htb]
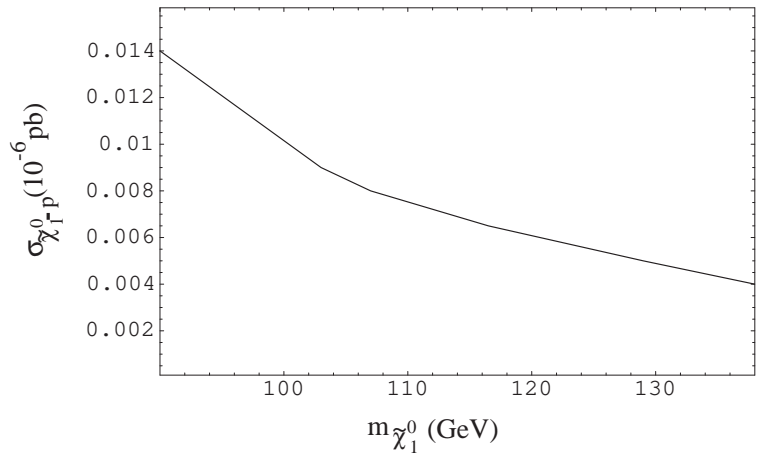

\centerline{ \DESepsf(aadslowlimit.epsf width 10 cm) }
\caption {\label{fig5}Minimum $\sigma_{\tilde\chi^0_1-p}$ for tan$\beta=3$ for 
$m_{1/2} < 345$ GeV[7].}
\end{figure}

In this region there is no coannihilation, and the smallest cross sections
occur at the smallest values of tan$\beta$. Fig.\ref{fig5} shows the minimium value of
the $\sigma_{\tilde\chi^0_1-p}$ cross section as a function of $m_{\tilde\chi^0_1}$
 for tan$\beta$=3. One sees that the
cross section
decreases with increasing $m_{\tilde\chi^0_1}$ as expected, and in this domain we have
\begin{equation}
         \sigma_{\tilde\chi^0_1-p}\stackrel{>}{\sim} 4\times 10^{-9} {\rm
	 pb}\,\, \mbox{ for }
\, m_{\tilde\chi^0_1} \leq 140 \mbox{ GeV}.
\end{equation}which would be accessible to planned future experiments (such as GENIUS).

\subsubsection{Coannihilation region ($m_{\tilde\chi^0_1}\stackrel{>}{\sim}150$
 GeV)}
Coannihilation in the early universe occurs when a second SUSY particle
becomes nearly degenerate with the neutralino LSP, and hence increase the
annihilation cross section. This effect is significant in mSUGRA due to two 
``accidents":
(1) The ${\tilde\chi^0_1}$ is a Majorana spinor and so its early universe annihilation cross
section $\sigma_{\tilde\chi^0_1-\tilde\chi^0_1}$ 
is supressed relative to e.g. R-slepton ($\tilde l_R$)
annihilations:
\begin{equation}
      \sigma_{\tilde\chi^0_1-\tilde\chi^0_1}\simeq (1/10) 
      (\sigma_{\tilde\chi^0_1-\tilde l_R},\, \sigma_{\tilde l_R\tilde l_R})
\end{equation}

(2) There is an accidental near degeneracy between $l_R$ and ${\tilde\chi^0_1}$ in a small
region of parameter space. To see this, one may look at low tan$\beta$, where
one has the analytic formulae for the selectron and neutralino masses of
\begin{equation}
m_{e_R}^2 = m_0^2  + (6/5)(\alpha_G/4\pi)f_1m_{1/2}^2 -\sin^2\theta_W M_W^2
\cos 2\beta
\end{equation}
\begin{equation}
    m_{\tilde\chi^0_1}\simeq (\alpha_1/\alpha_G)m_{1/2}
\end{equation}
where $f_1=(1/\beta_1)[1-(1/(1+\beta_1 t)]$ and $\beta_1$ is the $U(1)$ beta function
and $t=2\mbox{ln}[M_G/M_Z]$. Numerically, Eqs.(13) and (14) give
\begin{equation}
              m_{eR}^2 \simeq m_0^2 \,+\, 0.15 m_{1/2}^2\, +\, (40\, \rm GeV)^2
\end{equation}

 \begin{equation} 
 m_{\tilde\chi^0_1}^2\simeq 0.16 m_{1/2}^2
\end{equation}
One sees that for small $m_0$, the $\tilde e_R$ can become degenerate with the 
$\tilde\chi^0_1$, and
as $m_{1/2}$ increases, $m_0$ correspondingly increases to maintain the region of
near degeneracy. Thus one has a corridor in the $m_0-m_{1/2}$ plane of increasing
$m_0$ and $m_{1/2}$ where coannihilation effects can occur, extending the region
where Eq(4) can be satisfied. The importance of this effect has been
stressed in [14], where the analysis has been carried out for low and
intermediate tan$\beta$, an example of
\begin{figure}[htb]
\centerline{ \DESepsf(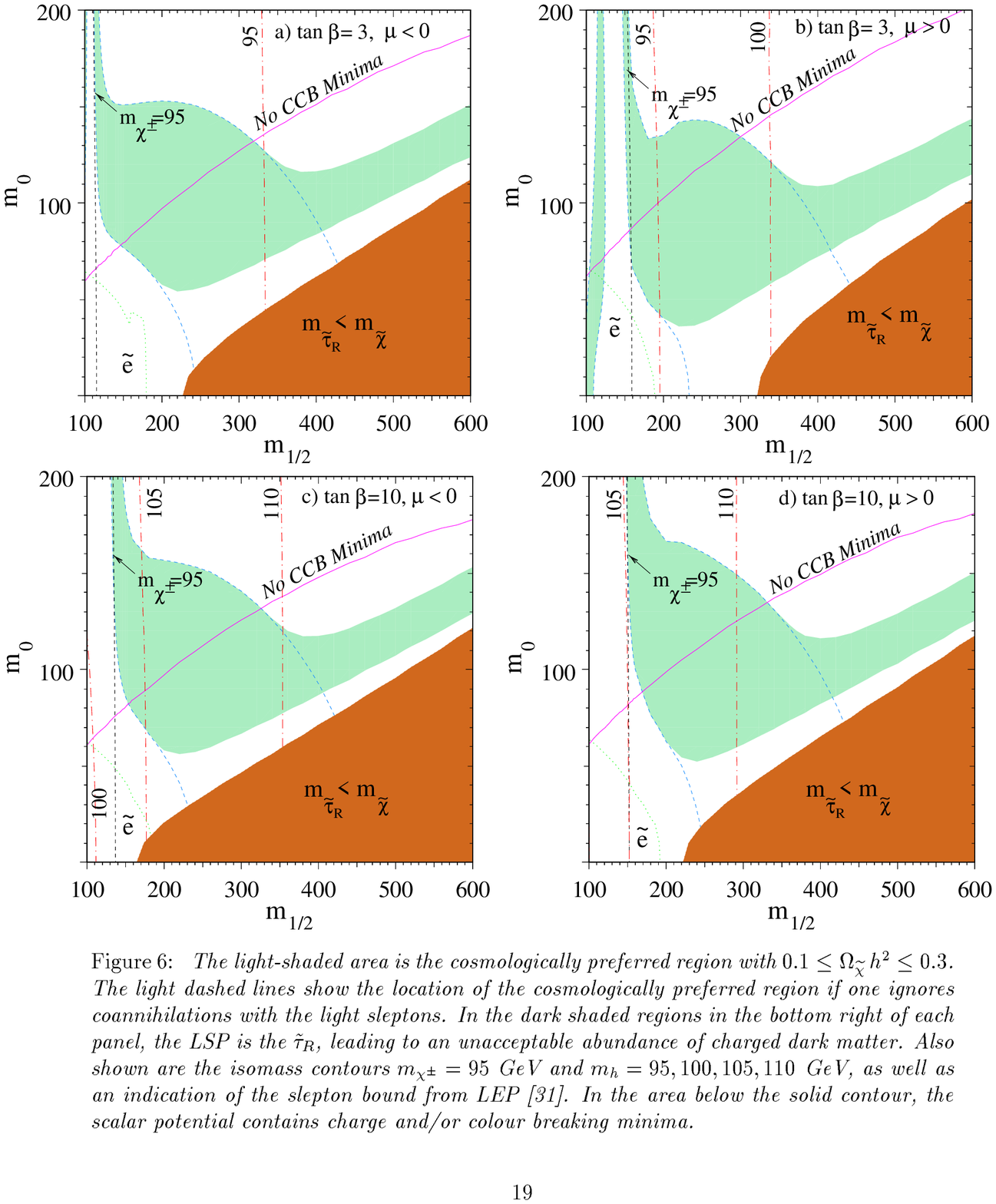 width 10 cm) }
\caption {\label{fig6} Fig. 6 of Ellis etal[14] }
\end{figure}
which is shown in Fig.\ref{fig6}. One sees that the coannihilation effect begins at
$m_{1/2}\stackrel{>}{\sim}400$ GeV (i.e. $m_{\tilde\chi^0_1}\stackrel{>}{\sim}150$GeV).
For large tan$\beta$, the sitiuation is more complicated as L-R mixing in the
slepton mass matrices becomes important, particularly for the third
generation, and generally the light $\tilde\tau_1$ is the lightest slepton,
considerably lighter than the other sleptons.
We consider first the case where $\mu > 0$. (We use the Isajet sign convention for $\mu$).
Fig.\ref{fig7} shows the allowed region in the $m_0-m_{1/2}$ plane where Eq.(4) is
satisfied for tan$\beta=$ 40. One sees that there is a significant $A_0$
\begin{figure}[htb]
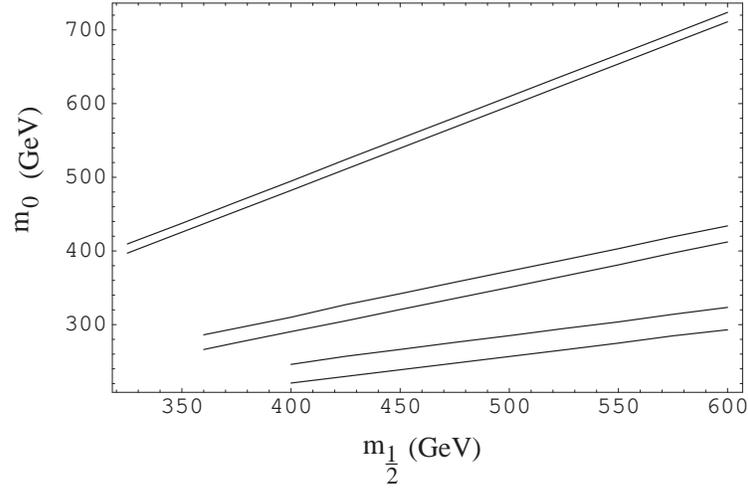

\centerline{ \DESepsf(aadcoan40.epsf width 10 cm) }
\caption {\label{fig7}Allowed corridors in $m_0-m_{1/2}$ plane satisfying the relic density
constraint Eq(4) for tan$\beta$=40, $\mu>$0 and (from bottom to top) 
$A_0=m_{1/2}$,
$2m_{1/2}$, $4m_{1/2}$ [7].}
\end{figure}
dependence, with larger $A_0$ allowing for larger $m_0$. In general the thickness
of the corridor is $\delta m_0\simeq25$ GeV. There is no longer any non
coannihilation region left, as the corridors terminate at $m_{1/2}$ above the
non coannihilation domain (for large tan$\beta$). (The termination is due to
the $m_h$ and $b\rightarrow s\gamma$ constraints.) The minimum value of $m_{1/2}$ decreases
with increasing $A_0$, as does the thickness of the allowed corridor. Thus for
very large $A_0$, the existence of these corridors eventually becomes a fine
tuning.
Since larger values of $A_0$ allow for larger values of $m_0$, one expects the
$\sigma_{\tilde\chi^0_1-p}$ cross section to decrease with $A_0$. This is illustrated in
 Fig.\ref{fig8} where
$\sigma_{\tilde\chi^0_1-p}$ is given as a function of $m_{1/2}$ for $\tan
\beta = 40$ for two values of $A_0$.
Thus one expects
\begin{figure}[htb]
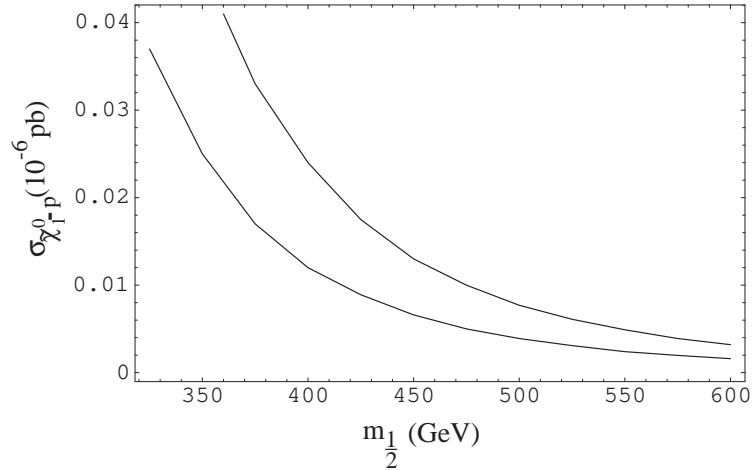

\centerline{ \DESepsf(aadcoan40A.epsf width 10 cm) }
\caption {\label{fig8}$\sigma_{\tilde\chi^0_1-p}$ as a function of $m_{1/2}$
 for tan$\beta=$40, $\mu >0$, $A_0=2m_{1/2}$
(upper curve) and $A_0=4m_{1/2}$ (lower curve)[7].
 }
\end{figure}
the minimum detection cross section to occur at largest $A_0$ and smallest
tan$\beta$. This is illustrated in Fig.\ref{fig9}. where the cross section is plotted
for $A_0=4m_{1/2}$, $\mu>$0, tan$\beta=40$ and tan$\beta=3$. 
Because the higher tan$\beta$
allows $m_0$ to become
\begin{figure}[htb]
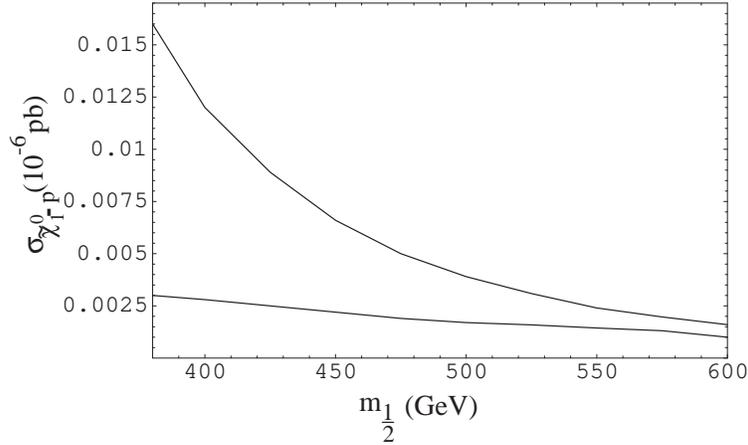

\centerline{ \DESepsf(aadcoan403.epsf width 10 cm) }
\caption {\label{fig9}$\sigma_{\tilde\chi^0_1-p}$ as a function of $m_{1/2}$ 
for 
$A_0=4m_{1/2}$, 
$\mu>0$, $\tan\beta=40$ (upper
curve),$\tan\beta=3$ (lower curve)[7].}
\end{figure}
larger (compare with Fig.\ref{fig6}) which also reduces the cross section, the
tan$\beta$  dependence is mostly neutralized for large $m_{1/2}$. One has, however,
the lower bound on the cross section of
\begin{equation}
       \sigma_{\tilde\chi^0_1-p}\stackrel{>}{\sim} 1\times10^{-9}\,{\rm pb},\,\,
       {\rm for}\,\,
        m_{1/2} <600\,\mbox{GeV }\, (m_{\tilde\chi^0_1}<240\,{\rm GeV}),\, \mu>0
\end{equation}
which should still be accesible to the proposed future detectors.

\begin{figure}[htb]
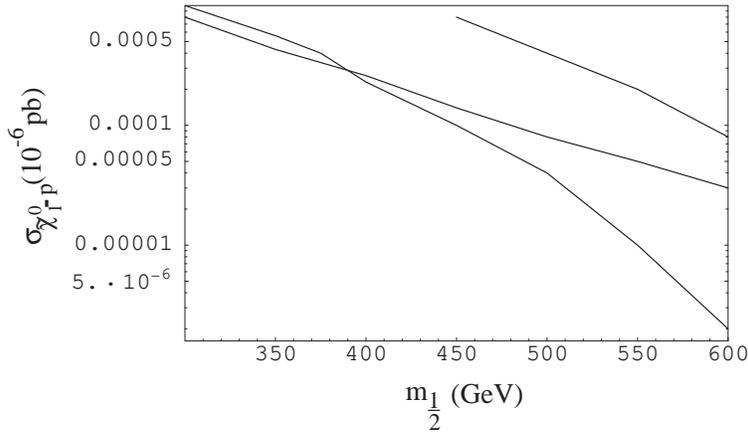

\centerline{ \DESepsf( aadcoan51020.epsf width 10 cm) }
\caption {\label{fig10} $\sigma_{\tilde\chi^0_1-p}$ for 
$\mu<0$, tan$\beta=$ 20, 5, 10 (descending order on the right)[7].}
\end{figure}

We next turn to the case of $\mu <0$.
As pointed out in [15], at low and intermediate tan$\beta$, an accidental
cancellation can occur in part of the parameter space in the coannihilation
region which can greatly reduce $\sigma_{\tilde\chi^0_1-p}$. We investigate here whether this
cancellation continues to occur in the high tan$\beta$ region. What occurs is
seen in Fig.\ref{fig10} where $\sigma_{\tilde\chi^0_1-p}$ is shown for tan$\beta=$20, 5 and 10 (in
descending order). One sees that
the cross section decreases  between tan$\beta$ = 5 and tan$\beta$ = 10, but then
rises again at higher tan$\beta$. Thus one has that
\begin{equation}
 \sigma_{\tilde\chi^0_1-p}< 10^{-10}\mbox{pb } \,{\rm for}
 \,\, 4\stackrel{<}{\sim}\tan\beta\stackrel{<}{\sim}20;\,\, m_{1/2}>450\,{\rm GeV}
  (m_{\tilde g}\stackrel{>}{\sim}\,1.1\mbox{ TeV});\, \mu>0
\end{equation}
and the minimum cross section occurs at tan$\beta\cong10$:
\begin{equation}
         (\sigma_{\tilde\chi^0_1-p})_{\rm min}
	  \simeq1\times10^{-12}\,{\rm pb}\,\, {\rm at}\,\,
	   \tan\beta=10,\, m_{1/2}\simeq 600\, {\rm GeV}
\end{equation}

Thus in this region of parameter space the proposed future detectors would
not be able to detect mSUGRA ${\tilde\chi^0_1}$ wimps. However, the absence of detection of
halo wimps would then imply that squarks and gluinos should lie above 1 TeV,
but at masses still acessible to the LHC. Also then, mSUGRA would require
that tan$\beta$ would be in the restricted range given in Eq(18), and $\mu$ be
negative. This would allow a number of cross checks on the validity of the
mSUGRA model.

\section{Nonuniversal SUGRA Models}

In most discussions of SUGRA models with nonuniversal soft breaking terms,
the universality of the the soft breaking masses at $M_G$ of the first two
generations of squarks and sleptons is maintained to suppress flavor
changing neutral currents. However, one may allow both the Higgs masses and
the third generation squark and slepton masses to become nonuniversal at
$M_G$. One can parameterize this situation at $M_G$ as follows:
\begin{eqnarray} m_{H_{1}}^{\
2}&=&m_{0}^{2}(1+\delta_{1}); 
\quad m_{H_{2}}^{\ 2}=m_{0}^{2}(1+ \delta_{2});\\\nonumber m_{q_{L}}^{\
2}&=&m_{0}^{2}(1+\delta_{3}); \quad m_{t_{R}}^{\ 2}=m_{0}^{2}(1+\delta_{4});
\quad m_{\tau_{R}}^{\ 2}=m_{0}^{2}(1+\delta_{5}); 	\\\nonumber m_{b_{R}}^{\
2}&=&m_{0}^{2}(1+\delta_{6}); \quad m_{l_{L}}^{\ 2}=m_{0}^{2}(1+\delta_{7}).
\label{eq18}
\end{eqnarray}
where $q_{L}\equiv (\tilde t_L, \tilde b_L)$ squarks, $l_{L}\equiv (\tilde \nu_\tau, \tilde \tau_L)$ sleptons,
 etc. and $m_0$ is
the universal mass for the first two generations of squarks and sleptons.
The $\delta_i$ are the deviations from universality (and if one were to impose
SU(5) or SO(10) symmetry one would have $\delta_3$=$\delta_4$=
$\delta_5$, and
$\delta_6$=$\delta_7$.) In the following we limit the $\delta_i$ to obey:
\begin{equation}
                 -1 \leq \delta_i \leq +1
\end{equation}
and maintain gauge coupling constant unification and gaugino mass
unification at $M_G$.

\begin{figure}[htb]
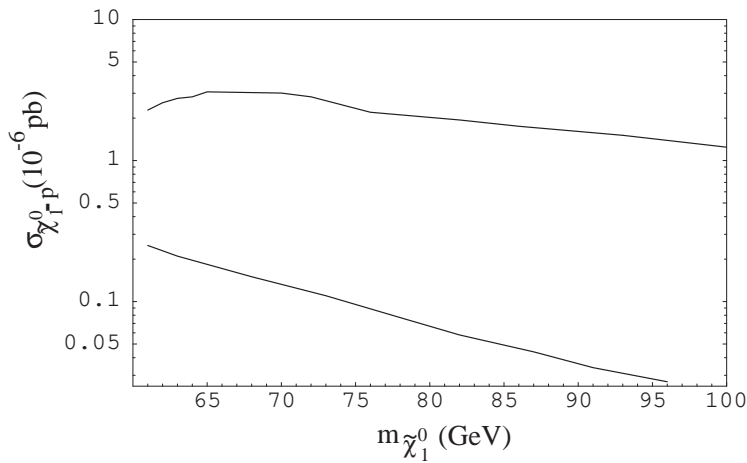

\centerline{ \DESepsf(aads7uni7nonuni.epsf width 10 cm) }
\caption {\label{fig11} Maximum $\sigma_{\tilde\chi^0_1-p}$ for tan$\beta=7$, 
$\mu>$0 for nonuniversal model,
$\delta_3$, $\delta_4$, $\delta_1<0$, $\delta_2>$0 (upper curve), 
and universal model (lower
curve)[13].}
\end{figure}

\begin{figure}[htb]
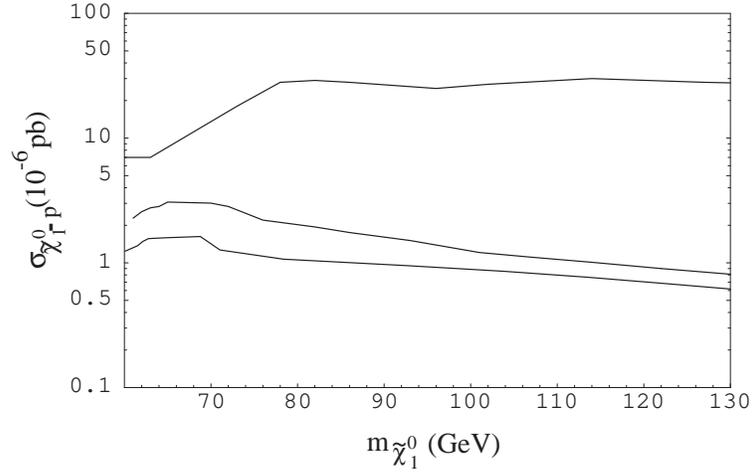

\centerline{ \DESepsf(aads5715nonuni.epsf width 10 cm) }
\caption {\label{fig12} Maximum $\sigma_{\tilde\chi^0_1-p}$ for nonuniversal SUGRA
models for tan$\beta=$5,7 and
15 (in ascending order)[13].}
\end{figure}

\begin{figure}[htb]
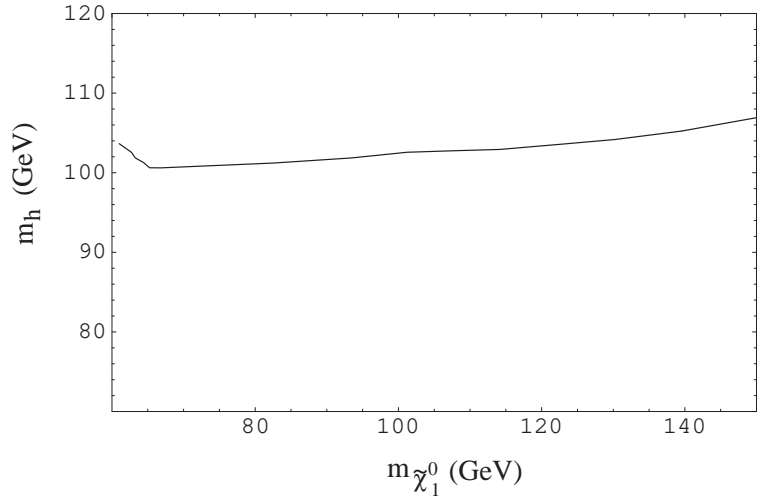

\centerline{ \DESepsf(aads7higgs1.epsf width 10 cm) }
\bigskip
\bigskip
\caption {\label{fig13} $m_h$ as a function of $m_{\tilde\chi^0_1}$ for tan$\beta=$7 when 
$\sigma_{\tilde\chi^0_1-p}$ takes on the
maximum value of Fig.\ref{fig12}[13].}
\end{figure}

While there are a large numbers of new parameters, one can get an
understanding of what effect they produce from the following. The
neutralino ${\tilde\chi^0_1}$ is a mixture of gaugino (mostly bino) and higgsino parts:
\begin{equation}
      {\tilde\chi^0_1}= \alpha \tilde W_3 + \beta \tilde B + \gamma \tilde H_1 +
       \delta \tilde H_2
\end{equation}
The dominant spin independent $\sigma_{\tilde\chi^0_1-p}$ cross section is proportional to the
interference between the gaugino and higgino amplitudes, and this
interference is largely governed by the size of $\mu^2$. As $\mu^2$ decreases,
the interference increases, and hence $\sigma_{\tilde\chi^0_1-p}$ increases. Radiative
breaking of $SU(2)\times U(1)$ determines the value of $\mu^2$ at the electroweak
scale. To see the general nature of the effects of nonuniverality, we
consider low and intermediate tan$\beta$ where an analytic form exists for
$\mu^2$ (see e.g. Arnowitt and Nath, Ref[2]):
\begin{eqnarray}
\mu^2&=&{t^2\over{t^2-1}}[({{1-3 D_0}\over 2}-{1\over
t^2})+({{1-D_0}\over2}(\delta_3+\delta_4)-
{{1+D_0}\over2}\delta_2+{\delta_1\over
t^2})]m_0^2\\\nonumber&+& {\rm {universal \, parts +loop \, corrections}}.
\end{eqnarray}
Here $t=\tan\beta$ and $D_0\cong1-(m_t/200\,{\rm GeV}\, \sin\beta)^2 \leq 0.2$ 
(Note that the
Higgs and squark nonuniversalties enter coherrently, roughly in the
combination $\delta_3+\delta_4-\delta_2$.) We see from Eq.(22) that $\mu^2$ is
reduced, and hence $\sigma_{\tilde\chi^0_1-p}$ incresed for 
$\delta_3$, $\delta_4$, $\delta_1<0$,
$\delta_2>0$,
and $\mu^2$ is increased for $\delta_3$, $\delta_4$, 
$\delta_1>0$, $\delta_2<0$. Thus one can
get significantly larger cross sections in the nonuniversal models with the
first choice of signs for the $\delta_i$, and one can reduce the cross
sections (though not by such a large amount) with the second choice.
The above analytic results are illustrated in Fig.\ref{fig11}, where the maximum
$\sigma_{\tilde\chi^0_1-p}$ for the universal and nonuniversal models are plotted for
tan$\beta=7$. One sees that
one can increase the cross section by a factor of 10 to 100 by an
appropriate choice of signs. Thus current detectors can probe regions of
lower tan$\beta$ for nonuniversal models than for the universal one. The
allowed range can be seen from Fig.\ref{fig12}, where the maximum cross sections
are plotted for tan$\beta =$ 5,7,and 15.
Current detectors with sensitivity of Eq.(1) thus can probe parts of the
parameter space with tan$\beta \stackrel{>}{\sim} 4$, and from the tan$\beta=$ 15 curve, 
we see
that parts of the high tan$\beta$ part of the parameter space has already been
eliminated. However, the very low tan$\beta$ values are on the edge of being
eliminated by the LEP constraint on the light Higgs mass. Thus Fig.\ref{fig13}
shows that $m_h$ is quite small if
$m_{\tilde\chi^0_1}$ is light, and one would have to raise the lower bound on 
tan$\beta$ as LEP
raises the lower bound on $m_h$.

As in mSUGRA, the minimum cross sections occur for the largest $m_{1/2}$ and
smallest tan$\beta$, and so they occur in the coannihilation region. We
consider here only the case where the Higgs masses are nonuniversal i.e.
$\delta_{1,2} \neq0$ (the other $\delta_i$ set to zero). Results then are similar to
the mSUGRA case. For $\mu>0$ we find
\begin{equation}
\sigma_{\tilde\chi^0_1-p}\stackrel{>}{\sim} 1\times 10^{-9} {\rm pb};\,\,{\rm
for}\,\,
 m_{1/2} \leq 600\, {\rm GeV},\,\mu>0
\end{equation} 
For $\mu<$0 one again can get a cancelation reducing the cross section to a
minimum near tan$\beta=$ 10:
\begin{equation}
        \sigma_{\tilde\chi^0_1-p}\stackrel{>}{\sim} 1\times 10^{-12} {\rm pb} \,
	\,{\rm at}\,\,
 m_{1/2} =600\,{\rm GeV},\, \tan\beta\cong10,\, \mu<0
\end{equation}

\section{Conclusions}
We have examined here the predictions of several SUGRA models which possess
gauge coupling constant unification at $M_G\cong2\times10^{16}$ GeV, to see what parts
of the SUSY parameter space are accessible to current detectors obeying
Eq.(1), and what will be accessible to future detectors with the
sensitivity of Eq.(2). For the minimal SUGRA model with universal soft
breaking parameters, mSUGRA, current detectors are scanning parts of the
parameter space where tan$\beta\stackrel{>}{\sim} 25$, 
$m_{\tilde\chi^0_1}\stackrel{<}{\sim} 90$
GeV and $\Omega h^2\stackrel{<}{\sim}0.1$. In
addition, the light Higgs obeys $m_h\stackrel{<}{\sim} 120$ GeV, and hence possibly would be
accessible to RUN2 at the Tevatron. For nonuniversal models, where one
allows the Higgs and third generation squark and slepton softbreaking
masses to be nonuniversal, the neutralino-proton cross section can be
significantly increased, by a factor of 10-100, with an appropriate choice
of sign in the soft breaking deviations from universality. Thus current
detectors here could scan regions of parameter space as low as tan$\beta\simeq4$,
though in these regions $m_h$ is very light, and the the minimum allowed
tan$\beta$ may have to be raised as LEP raises the the bound on the Higgs
mass. However, the possibility of large cross section here has already
allowed current detectors to exclude parts of the high
tan$\beta$ region, e.g. when tan$\beta \stackrel{>}{\sim}$15.
How low SUGRA cross sections can lie is complicated by the existance of
coannihilation effects where the R-sleptons (particularly the $\tilde\tau_1$) can
become nearly degenerate with the neutralino. This allows the relic density
constraint Eq.(4) to be satisfied in a narrow rising corridor (about 25 GeV
wide in $m_0$) in the $m_0-m_{1/2}$ plane rising to relatively large $m_0$, and thus
reducing the size of the ${\tilde\chi^0_1}-p$ cross section. For large tan$\beta$, the effect
is sensitive to the value of $A_0$, the range of $m_0$ increasing with $A_0$. Thus
for $\mu >$0, one finds for $m_{1/2}=$600 GeV, the minimum cross section at e.g.
tan$\beta=40$, $A_0=4 m_{1/2}$ is almost the same as that at tan$\beta=3$, the increase
of the cross section due to the increase in tan$\beta$ being offset by the
decrease due to the allowed large value of $m_0$. One finds however, that for
$\mu>0$, at m$_{1/2}$ = 600 GeV ($m_{\tilde g}\stackrel{\sim}{=}1.5$ TeV), the minimum cross sections would still
be accessible to detectors with the sensitivity of Eq.(2).
The minimum cross sections for $\mu < 0$ is complicated by the possibility of
accidental cancelations in the scattering amplitudes, allowing the cross
section to sink below 10$^{-10}$ pb in certain regions of the parameter space.
Thus one finds for mSUGRA that $\sigma_{\tilde\chi^0_1-p }< 1\times 10^{-10}$ pb for
 $m_{1/2} \stackrel{>}{\sim}450$ GeV
($m_{\tilde g}\stackrel{>}{\sim}1.1$ TeV) when 
$4 \stackrel{<}{\sim}\tan\beta \stackrel{<}{\sim}$ 20, 
with a minmum cross section of $1\times
10^{-12}$ pb reached at $\tan\beta\cong$10. Similar results hold for the
nonuniversal models. In this domain SUGRA models imply that halo dark
matter would not be accessible to detectors with sensitivities of Eq.(2),
and that the gluino and squarks would be quite heavy. They would still
however be observable at the LHC which can detect gluinos with mass 
$m_{\tilde g}\stackrel{<}{\sim}$ 2.5 TeV [16].
\section{Acknowledgement}
This work was supported in part by National Science Foundation Grant No.
PHY-9722090.

\end{document}